\documentclass[twocolumn,pra,showpacs,floats,superscriptaddress]{revtex4}
\usepackage[dvips]{graphicx}
\usepackage{amsmath,bm}
\usepackage{psfrag}

\begin{document}

\title{Oscillatory and monotonic modes of longwave Marangoni convection in a thin film}

\author{S.~Shklyaev}
\affiliation{Department of Theoretical Physics, Perm State
University, 15 Bukirev St., Perm 614990, Russia}
\author{M.~Khenner}
\affiliation{Department of Mathematics, Western Kentucky University, Bowling Green, KY, 42101, USA}
\author{A.~A.~Alabuzhev}
\affiliation{Institute of the Continuous Media Mechanics, Ural
Branch of the Russian Academy of Sciences, Perm 614013, Russia}
\affiliation{Department of Theoretical Physics, Perm State
University, 15 Bukirev St., Perm 614990, Russia}


\begin{abstract}
We study longwave Marangoni convection in a layer heated from
below. Using the scaling $k = O(\sqrt{\rm Bi})$, where $k$ is the
wavenumber and ${\rm Bi}$ is the Biot number, we derive a set of
amplitude equations. Analysis of this set shows presence of
monotonic and oscillatory modes of instability. Oscillatory mode
has not been previously found for such direction of heating.
Studies of weakly nonlinear dynamics demonstrate that stable
steady and oscillatory patterns can be found near the stability
threshold.

\end{abstract}


\date{\today}

\maketitle

\underline{\textit{Introduction.}} Marangoni convection in a
liquid layer with upper free boundary is a classical problem in
the dynamics of thin films and in the pattern formation
\cite{books,reviews}. In the pioneer theoretical paper, Pearson
\cite{Pearson} analyzed the linear stability of the layer with a
nondeformable free surface. He considered two cases of thermal
boundary conditions at the substrate: the ideal and poor heat
conductivity, when either the temperature or the heat flux are
specified. In the latter case he found a monotonic longwave
instability mode for heating from below and zero Biot number
$\rm Bi$. For ${\rm Bi} \ll 1$ the critical wavenumber $k$ is
proportional to ${\rm Bi}^{1/4}$ \cite{books}. Many authors
extended the analysis in order to include the deformation of the
free surface. Review of analytical and numerical works can be
found in \cite{books}. In particular, several oscillatory modes
were revealed; these modes were reported only for heating from
above.

In the case of heating from below, a nonlinear analysis for
ideally conductive substrate was performed in Ref.~\cite{VanHook}:
it was shown that the subcritical bifurcation occurs and
instability with necessity results in film rupture. The
behavior of perturbations near the stability threshold was studied
in \cite{G_YV} for the case of a poorly conductive substrate.
Under assumption of large gravity, and, hence, small surface
deflection, the amplitude equation was derived and the subcritical
bifurcation was found.

In this paper we demonstrate the existence of a new
\textit{oscillatory} mode of longwave instability for the film
\textit{heated from below}. Using the scaling $k=O(\sqrt{\rm
Bi})$, which was first suggested in Ref.~\cite{Alla-05}, we derive
a set of amplitude equations. Linear stability analysis gives both
the monotonic and the oscillatory modes. Pattern selection near
the stability threshold clearly demonstrates that instability does
not necessarily lead to rupture and that both steady and
oscillatory regimes can be found experimentally within certain
domains of parameters.

\underline{\textit{Problem formulation.}} We consider a
three-dimensional thin liquid film of the unperturbed height $H_0$
on a planar horizontal substrate heated from below. The heat
conductivity of the solid is assumed small in comparison with the
one of the liquid, thus the constant vertical temperature gradient
$-A$ is prescribed at the substrate. (The Cartesian reference
frame is chosen such that the $x$ and $y$ axes are in the
substrate plane and the $z$ axis is normal to the substrate.)

The dimensionless boundary-value problem governing the fluid
dynamics reads:
\begin{subequations}
\label{base_eq}
\begin{eqnarray}
\frac{1}{P}\left({\bf v}_t + {\bf v}\cdot \nabla {\bf
v}\right)&=& - {\bf \nabla} p + \nabla ^2 {\bf v} -G {\bf e}_z,\\
T_t + {\bf v}\cdot \nabla T &=& \nabla ^2 T, \ \nabla \cdot {\bf
v} = 0,
\end{eqnarray}
\end{subequations}
\vspace{-8mm}
\begin{subequations}
\label{base_bcs}
\begin{eqnarray}
\label{bcs_velo}{\bf v} &=& 0, \ T_z=-1 \ {\rm at} \ z = 0, \\
\nonumber {\bf \Sigma}\cdot {\bf n} &=&\left(p-{\rm Ca}
K\right){\bf n}- M\nabla_\tau \left( T|_{z=h}\right),\
\nabla_n T=-{\rm Bi}\; T,\\
h_t &=& w-{\bf v} \cdot {\bf \nabla} h
 \ {\rm at} \ z = h(x,y,t).
\end{eqnarray}
\end{subequations}
Here, ${\bf v}=({\bf u},w)$ is the fluid velocity (where ${\bf u}$
is the velocity in the substrate plane and $w$ is the
$z$-component), $T$ is the temperature, $p$ is the pressure in the
liquid, $\bf \Sigma$ is the viscous stress tensor, $h$ is the
dimensionless height of the film, ${\bf e}_{z}$ is the unit vector
directed along the $z$ axis, ${\bf n}$ and $\bf \tau$ are the
normal and tangent unit vectors to the free surface, respectively,
$K$ is the mean curvature of the free surface. The dimensionless
parameters entering the above set of equations are the capillary
number, the Marangoni number, the Galileo number, the Biot number,
and the Prandtl number:
$$
{\rm Ca}=\frac{\sigma H_0}{\eta \chi}, \ M=-\frac{\sigma_T A
H_0^2}{\eta \chi}, \ G=\frac{g H_0^3}{\nu\chi}, \ {\rm
Bi}=\frac{qH_0}{\kappa},
$$
%
and $P=\nu/\chi$. Here $\sigma$ is the surface tension,
$\sigma_T\equiv d\sigma/dT$, $g$ is the acceleration of gravity, $q$ is the heat transfer rate,
$\kappa$ is the thermal conductivity, $\chi$ is the thermal
diffusivity, $\nu$ and $\eta $ are the kinematic and dynamics
viscosity of liquid, respectively.

%
%
Below we study
evolution of a large-scale convection
using the set of Eqs.~(\ref{base_eq}) and (\ref{base_bcs}).

\underline{\textit{Amplitude equations.}} We rescale the
coordinates and the time as follows:
\begin{equation} \label{rescaling}
X=\epsilon x, \ Y=\epsilon y, \ \tau=\epsilon^2 t,
\end{equation}
where $\epsilon \ll 1$ is the ratio of $H_0$ to a typical
horizontal lengthscale. The temperature field is represented as
$T=-z+{\rm Bi}^{-1}+\theta(X,Y,\tau)+O(\epsilon^2)$.

We assume large values of ${\rm Ca}$ and small values of $\rm Bi$,
\begin{equation}\label{Biot}
{\rm Ca}=\epsilon^{-2}C, \ {\rm Bi}=\epsilon^2 \beta.
\end{equation}
Thus we deal with the intermediate asymptotics between the
conventional longwave mode, ${\rm Bi}=O\left(\epsilon^4\right)$,
\cite{G_YV} and the case of finite $\rm Bi$ \cite{Pearson}. These
cases correspond to $\beta=0$ and $\beta\to\infty$, respectively.

Substituting the rescaled fields into Eqs.~(\ref{base_eq}) and
(\ref{base_bcs}) and applying the conventional technique of the
lubrication approximation (see \cite{reviews}), we arrive at
\begin{eqnarray}
\label{h_t} h_{\tau}&=&{\bf \nabla}\cdot\left[\frac{h^3}{3} {\bf
\nabla} \Pi + \frac{M h^2}{2}{\bf \nabla} \left(\theta-h\right)
\right] \equiv {\bf
\nabla \cdot j},\\
\nonumber h \theta_\tau&=&{\bf \nabla}\cdot \left(h{\bf
\nabla}\theta\right) -\frac{1}{2}(\nabla h)^2-\beta(\theta-h)+
{\bf j} \cdot {\bf \nabla}(\theta-h) \\
&&+ {\bf \nabla}\cdot \left[\frac{h^4}{8}{\bf \nabla}
\Pi+\frac{Mh^3}{6}{\bf \nabla}(\theta-h)\right].\label{T_t}
\end{eqnarray}
Here
$\Pi=Gh-C\nabla^2 h$ and $\nabla$ is a
two-dimensional gradient with respect to $X$ and $Y$.

Equations~(\ref{h_t}) and (\ref{T_t}) form a closed set of the
amplitude equations governing the nonlinear interaction
of two well-known longwave modes: the Pearson's mode ($h=1$)
\cite{Pearson} and the surface deformation-induced mode. (Note that the latter mode
with $\theta=const$ emerges only in the case of the conductive substrate
\cite{VanHook}.) Conductive state obviously corresponds to
$h=\theta=1$.

\underline{\textit{Linear stability analysis.}} Substituting the
perturbed fields $h=1+\xi$ and $\theta=1+\Theta$ into
Eqs.~(\ref{h_t}) and (\ref{T_t}), linearizing the equations for
perturbations about the equilibrium,
%
%
and representing the perturbation fields proportional to $\exp
\left(\lambda \tau+ikX\right)$, one arrives at

\begin{eqnarray}
\nonumber \lambda^2+\lambda\left[\beta+k^2\left(1+\frac{\tilde
G-M}{3}\right)\right]\\
+\frac{k^2}{3}\left(\beta+k^2\right)\tilde
G-\frac{Mk^4}{2}\left(1+\frac{\tilde
G}{72}\right)=0,\label{lambda}
\end{eqnarray}
where $\tilde G\equiv G+Ck^2$.
Equation (\ref{lambda}) possesses both real (monotonic
instability) and complex (oscillatory instability) solutions.

For
the {\em monotonic} mode $\lambda=0$ at the stability border, thus
the marginal stability curve is given by
\begin{equation}\label{M_mono}
M_{m}=\frac{48\left(\beta+k^2\right)\tilde G}{k^2\left(72+\tilde G\right)}.
\end{equation}
%
These marginal curves have a minimum at the finite values of $k$
only if
\begin{equation}\label{beta_lim}
\beta C<72,
\end{equation}
otherwise the minimal value, $M_c^{(m)}$, is achieved in the limit
$k\to \infty$, i.e. the longwave mode is not critical. Hereafter
we assume that the inequality (\ref{beta_lim}) holds; since the
limit $C=0$  is well studied \footnote{For $C=0$ (i.e., ${\rm Ca}$
is finite) the critical Marangoni number reduces to the
conventional value $48G/(G+72)$ \cite{G_YV}, which is approached
as $k\to \infty$. The same $M_c^{(m)}$ holds for $\beta=0$ as
well, but with zero critical wavenumber.}, for all computations we
set $C=1$ without loss of generality \footnote{This can be
achieved by the rescaling of Eqs.~(\ref{h_t}) and (\ref{T_t}):
$(X,Y)\to \sqrt{C}(X,Y),\tau \to C \tau, \, \beta\to\beta/C$.}.
The critical wavenumber materializing the minimum of the marginal
stability curve, Eq.~(\ref{M_mono}), is
$$
\left(k_c^{(m)}\right)^2=\frac{\beta C G+\sqrt{72\beta C
G\left(G+72-\beta C\right)}}{C\left(72-\beta C\right)}.
$$
%

%
\begin{figure}
\includegraphics[width=8.5 cm]{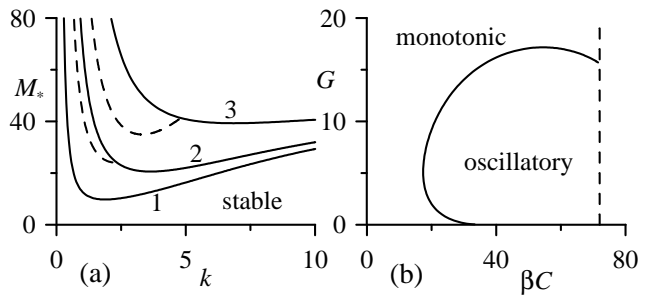}
\caption{(a): Marginal stability curves $M_{*}(k)$ for $G=10$:
solid lines correspond to the monotonic mode, dashed ones -- to
the oscillatory mode; $\beta=1, \, 10,\, 40$ for lines 1, 2, and
3, respectively.
(b): The domain of oscillatory instability. The dashed vertical
line marks the boundary of the longwave instability,
Eq.~(\ref{beta_lim}).} \label{fig:Mk}
\end{figure}

For the {\em oscillatory} mode the marginal stability curve is
determined by the expression
\begin{equation}\label{M_osc}
M_{o}=3+\tilde G+\frac{3\beta}{k^2}.
\end{equation}
The imaginary part of the growth rate for neutral perturbations is
\begin{equation}
\lambda_i\equiv {\rm Im} (\lambda)
=\frac{k^2}{12}\sqrt{(72+\tilde G)\left(M_{m}-M_{o}\right)},
\end{equation}
i.e. the oscillatory mode is present only at $M_{o}(k)<M_{m}(k)$.

Minimization of the Marangoni number with respect to $k$ gives
\begin{equation}
M_{c}^{(o)}=3+G+2\sqrt{3\beta C}, \
k_c^{(o)}=\left(\frac{3\beta}{C}\right)^{1/4}. \label{Mc_osc}
\end{equation}

Examples of the marginal stability curves for these modes are
shown in Fig.~\ref{fig:Mk}(a). Domains of monotonic and
oscillatory instability are demonstrated in Fig.~\ref{fig:Mk}(b).
It is clear that the oscillatory mode is critical for $\beta C >
17.4$ and $G<17.2$. Take, for instance, a layer of water of
thickness $H_0=10^{-3} {\rm cm}$.  Then $G\approx 0.1$, ${\rm
Ca}\approx 10^4$ and $\rm Bi$ has to be approximately $10^{-3}$ in
order to provide the required value of $\beta C$; this value seems
achievable in experiments.

Equations~(\ref{M_osc})-(\ref{Mc_osc}) indicate why the
oscillatory mode has not been found earlier. As we have emphasized
above, all previous studies deal with either $\tilde G \gg 1$
\cite{Pearson}, or $\beta=0$ \cite{G_YV}, or $C=0$ \cite{Alla-05}.
In these cases the oscillatory mode does not exist.

\underline{\textit{Weakly nonlinear analysis. Monotonic mode.}}
Here we study the nonlinear dynamics of perturbations at small
supercriticality, $M-M_c^{(m)}\approx 0$, see
Ref.~\cite{Hoyle-book}. To this end, we represent the primary part
of the small perturbation of $h$ in the form:

\begin{figure}
\includegraphics[width=8.5 cm]{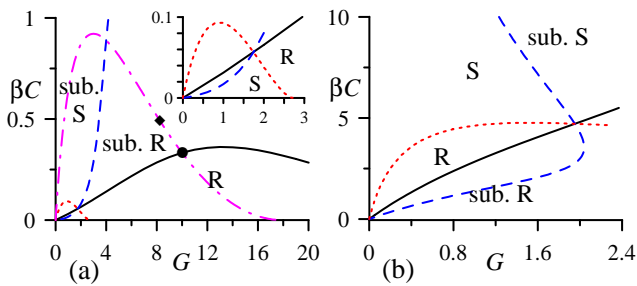}
\caption{(Color online). Pattern selection for the monotonic mode.
(a) and (b) -- the domains of stability for Rolls (marked with
``R'') and Squares (``S'') on the {\em square} lattice. Solid
(dashed) lines separate between supercritical and subcritical
branching for Rolls (Squares). The latter domains are marked by
``sub. R'' (``sub. S''). Dotted lines separate domains of
stability for Rolls and Squares. Dashed-dotted line in panel (a)
is the locus of points $N=0$; in the vicinity of this curve
Eq.~(\ref{hex3}) holds. Diamond (circle) shows the threshold value
$G_1$ ($G_2$) for pattern selection on the {\em hexagonal}
lattice.} \label{fig:domains}
\end{figure}

%
\begin{equation}
\xi=\sum_{j=1}^{n} A_{j} \exp\left(i{\bf k}_j\cdot {\bf
R}\right)+c.c.
\label{mono_lin}
\end{equation}
where $c.c.$ denotes complex conjugate terms and $k_j=k_c^{(m)}$.
(The primary part of $\Theta$ is expressed in terms of
$\xi$.) The amplitudes $A_j$ are functions of a slow time. For
{\em square} ($n=2$) and {\em hexagonal} ($n=3$) lattices, the
wavevectors are
\begin{eqnarray}
\label{k_sq} {\bf k}_1&=&k_c(1,0), \ {\bf k}_2=k_c(0,1)\\
\label{k_hex}   {\rm and} \ {\bf k}_1&=&k_c(1,0), \ {\bf
k}_{2,3}=\frac{1}{2}k_c(-1,\pm \sqrt{3}),
\end{eqnarray}
respectively.

For {\em square lattice}, the amplitude equations read
\begin{eqnarray}
\dot A_j&=&\left(\gamma -K_0|A_j|^2-K_1S_A\right)A_j, \ j=1,2,
\end{eqnarray}
where $S_A = \sum_1^n |A_l|^2$. Here the dot denotes the
derivatives with respect to the slow time, and $\gamma\sim
M-M_c^{(m)}$ is the real growth rate. The Landau constants, $K_0$
and $K_1$ are real; they are cumbersome and thus are not presented
here. Results of the numerical calculations are shown in
Fig.~\ref{fig:domains}. One can readily see that supercritical
branching occurs only in two domains of parameters. These domains
are situated either at rather small values of $\beta C$,
Fig.~\ref{fig:domains}(a), or at sufficiently small $G$, Fig.
\ref{fig:domains}(b). In the former case Rolls are selected
everywhere except for a very small region shown in the inset. In
the latter case Squares are selected everywhere excluding the
small region where Rolls are stable.

For {\em hexagonal lattice}, the resonant quadratic interaction results
in the following amplitude equation:
%
%

%
\begin{equation}\label{hex3}
\dot A_1=\gamma A_1-N A_2^*A_3^*
-\left(K_0|A_1|^2+K_1S_A\right)A_1,
\end{equation}
and a similar equations for $A_{2,3}$. (Hereafter the asterisk
denotes the complex-conjugate terms.) Generally speaking, the
quadratic term prevails over cubic ones, which leads to
subcritical excitation of the hexagonal patterns through a
transcritical bifurcation \cite{Hoyle-book}. However, $N=0$ at the
dashed-dotted line shown in Fig.~\ref{fig:domains}(a) and in the
vicinity of this line Eq.~(\ref{hex3}) becomes appropriate.

Among the variety of possible patterns \cite{Hoyle-book}, three
are important. They are Rolls with $A_1\neq 0, \, A_2=A_3=0$ and
two types of Hexagons with $A_1=A_2=A_3\equiv A$: $H^+$ for $A>0$
and $H^-$ in the opposite case. In the former case the flow is
upward in the center of the convective cell, whereas in the latter
case it is downward.

Pattern selection on a hexagonal lattice is shown in
Fig.~\ref{fig:domains}(a). At $G<G_1\approx 8.20$ there are no
stable solutions; the subcritical bifurcation occurs for Rolls and
one branch of Hexagons (either $H^-$ below or $H^+$ above the
dashed-dotted line). At $G_1<G<G_2=10$ Rolls are still subcritical
and unstable; stable Hexagons emerge only within the finite
interval of supercriticality. Finally, at $G>G_2$, $H^-(H^+)$ is
stable within the interval of supercriticality, whereas Rolls
become stable when $M-M_c^{(m)}$ increases.

To finalize the discussion of steady patterns, we briefly discuss
the competition of patterns on the square and hexagonal lattices.
It is clear that at the finite values of $N$, Hexagons emerge
subcritically and no stable patterns can be found near the
stability threshold. Therefore, weakly nonlinear analysis provides
stable patterns only near the dashed-dotted curve shown in
Fig.~\ref{fig:domains}(a), where the competition between Hexagons
and Rolls occurs.

\underline{\textit{Weakly nonlinear analysis. Oscillatory mode.}}
For the oscillatory mode the solution is presented in the form
\begin{equation}\label{osc_lin}
\xi=\sum_{j=1}^n \left(A_{j} e^ {i{\bf k}_j\cdot {\bf
R}}+B_je^{-i{\bf k}_j\cdot {\bf R}} \right)e^{i\lambda_i\tau}+c.c.
\end{equation}
Note that the pair $(A_j,B_j)$ corresponds to counter-propagating
waves, which must be taken into account separately. The
wavevectors for the square and hexagonal lattices are given by
Eqs.~(\ref{k_sq}) and (\ref{k_hex}), respectively.

For {\em square lattice}, the equation governing
the dynamics of the amplitudes $A_j$ reads:
\begin{eqnarray}
\nonumber \dot A_j&=&\left[\gamma
-K_0|A_j|^2-K_1|B_j|^2-K_2\left(S_A+S_B\right)\right]A_j\\
&&-K_4B_j^*S_{AB}, \ j=1,2, \label{sq_osc}
\end{eqnarray}
where $S_B = \sum_1^n|B_l|^2$, $S_{AB}= \sum_1^nA_lB_l$. A similar
pair of equations for $B_j$ is obtained from Eqs.~(\ref{sq_osc})
by replacement $A_j\leftrightarrow B_j$. The Landau coefficients
$K_l$ ($l=0,1,2,4$) as well as the growth rate $\gamma$ are now
complex-valued.

\begin{figure}
\includegraphics[width=8.5 cm]{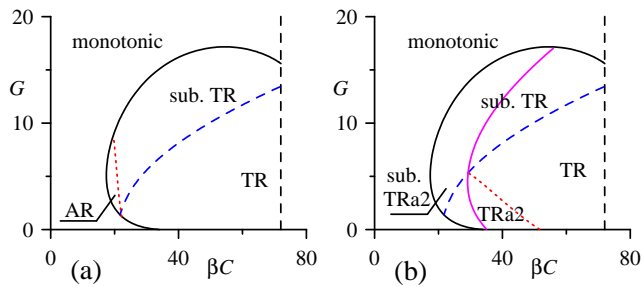}
\caption{(Color online). Pattern selection for the oscillatory
convection. (a) -- square lattice: Domains of stability for TR
(below the dashed line) and AR (to the left of the dotted line).
Above the dashed line TR bifurcate subcritically. (b) -- hexagonal
lattice: Domains of stability for TR (below the dashed line and to
the right of the dotted line) and TRa2 (between the dotted and the
solid line) are marked by ``TR" and ``TRa2'', respectively. Above
the dashed line TR bifurcate subcritically, to the left of the
solid line TRa2 are subcritical. } \label{fig:osc_sel}
\end{figure}

Equations (\ref{sq_osc}) were studied in details in
Ref.~\cite{Silber-Knobloch}. Using the results of that
paper, we found that Traveling Rolls (TR), $A_1\neq 0, \,
A_2=B_{1,2}=0$ can branch either supercritically or subcritically
[see Fig.~\ref{fig:osc_sel}(a)], whereas the remaining patterns
emerge through the direct Hopf bifurcation; TR are selected in the
domain of supercritical excitation.
Alternating Rolls are stable within the small area marked by ``AR''; here depending on the initial
condition the system either approaches AR or demonstrates the
infinite growth of TR.

For {\em hexagonal lattice}, the amplitude equation governing the
dynamics of the complex amplitudes $A_j$ reads:
\begin{eqnarray}
\nonumber \dot A_j&=&\left[\gamma
-K_0|A_j|^2-K_1|B_j|^2-K_2S_A-K_3S_B\right]A_j\\
&&-K_4B_j^*S_{AB}, \ j=1,2,3. \label{hex_osc}
\end{eqnarray}
Three similar equations are obtained from Eqs.~(\ref{hex_osc}) by
a replacement $A_j\leftrightarrow B_j$.

Analysis of the Hopf bifurcation for the above set of equations
was performed in Ref.~\cite{Roberts}, where eleven limit cycles
were found and studied.  Based on that paper, the results on
pattern selection are presented in Fig.~\ref{fig:osc_sel}(b). The
dashed line again separates direct and inverse Hopf bifurcations
for TR, it is obviously the same as in the panel (a). However, for
the hexagonal lattice, there appears a competition between TR and
Traveling Rectangles 2 (TRa2, $A_1=B_3\neq 0$, whereas all other
amplitudes vanish). The latter pattern is stable in the domain
marked by ``TRa2''. The entire domain of supercritical bifurcation
becomes smaller because TRa2 can bifurcate either supercritically
or subcritically.

Studying the competition between patterns on hexagonal and square
lattices, we found that the stability boundaries for both TR and
TRa2 are the same as shown in Fig.~\ref{fig:osc_sel}(b), whereas
stability domain for AR nearly disappears.

\underline{\textit{Conclusions.}} We studied the longwave
Marangoni convection in a liquid layer heated from below; the heat
flux at the substrate is specified. In such setup, an interaction
of two well-known monotonic modes of longwave instability, the
Pearson's mode and the surface deformation-induced mode, can
result in the emergence of a longwave oscillatory mode. However,
the
oscillatory mode has not been detected
in spite of extensive numerical, analytical, and experimental
studies \cite{books} since the publication of Pearson's paper. We succeed in such
analysis and point out the domain of parameters where the
oscillatory mode exists, which can be reached in experiments.

Moreover, we point out the domains of parameters where the
convection emerges supercritically and hence either stationary or
oscillatory terminal state with distorted surface is stable. This
result is also very unusual, since  only
subcritical branching was found in the previous studies \cite{VanHook,G_YV}.



\underline{\textit{Acknowledgments.}} We are grateful to
A.~A.~Nepomnyashchy and A.~Oron for the fruitful discussions. S.S.
and A.A. are partially supported by joint grants of the Israel
Ministry of Sciences (Grant 3-5799) and Russian Foundation for
Basic Research (Grant 09-01-92472). M.K. acknowledges the support
of WKU Faculty Scholarship Council via grants 10-7016 and 10-7054.

\end{document}